\documentclass{article}
\usepackage{spconf,amsmath,graphicx,hyperref}

\usepackage{microtype}

\usepackage{booktabs}
\usepackage[table]{xcolor}
\definecolor{wenetspeechmin}{RGB}{226,242,226}

\makeatletter
\renewcommand\section{\@startsection{section}{1}{\z@}%
  {-2.56ex \@plus -.3ex \@minus -.2ex}{1.3ex \@plus .1ex}{\normalfont\normalsize\bfseries}}
\renewcommand\subsection{\@startsection{subsection}{2}{\z@}%
  {-2.36ex \@plus -.3ex \@minus -.2ex}{0.9ex \@plus .1ex}{\normalfont\normalsize\bfseries}}
\renewcommand\subsubsection{\@startsection{subsubsection}{3}{\z@}%
  {-1.6ex \@plus -.3ex \@minus -.2ex}{0.6ex \@plus .1ex}{\normalfont\normalsize\itshape}}
\long\def\@makecaption#1#2{%
  \vskip 3pt
  \setbox\@tempboxa\hbox{#1. #2}%
  \ifdim \wd\@tempboxa >\hsize #1. #2\par
  \else \hbox to\hsize{\hfil\box\@tempboxa\hfil}%
  \fi}
\let\spconfbibliography\thebibliography
\renewcommand{\thebibliography}[1]{%
  \spconfbibliography{#1}%
   \setlength{\itemsep}{1pt}%
  \setlength{\parsep}{0pt}}
\makeatother
\title{WenetSpeech-Min: A Large-Scale Minnan Speech Corpus with Dual Transcriptions for Dialectal Speech Processing}
\name{\begin{tabular}{c}
      Haoyu Zhang$^{1,*}$, Chunjiang He$^{1,*}$ , Hongtao Li$^{2}$, Zeyu Zhu$^{1}$, Qituan Shangguan$^{2}$, \\
      Chengyou Wang$^{1}$, Jingbin Hu$^{1}$, Ziyu Zhang$^{1}$, Bingshen Mu$^{1}$, Yanbo Wang$^{3}$\\
      Shuai Wang$^{2,4,8}$, Jinhui Ye$^{4}$, Chengdong Liang$^{4}$ ,Binbin Zhang$^{4}$, Pengcheng Zhu$^{4}$\\
      Chuang Ding$^{5}$, Qiangze Feng$^{6}$, Qingyang Hong$^{7}$, Liumeng Xue$^{2,8}$, Lei Xie$^{1,\dagger}$ 
      \end{tabular}
      \thanks{$^*$ Equal contribution. $^\dagger$ Corresponding author.}
      }   

\address{$^{1}$Audio, Speech and Language Processing Group (ASLP@NPU), Northwestern Polytechnical University\\
         $^{2}$School of Intelligence Science and Technology, Nanjing University\\
         $^{3}$University of New South Wales
         $^{4}$WeNet Open Source Community  $^{5}$Moonstep AI\\
         $^{6}$Nexdata   $^{7}$School of Informatics, Xiamen University \\
         $^{8}$State Key Laboratory of Novel Software Technology, Nanjing University}

\begin{document}
\ninept
\maketitle
\begin{abstract}
Progress in dialectal speech technology is hindered by the scarcity of large-scale, real-world corpora. For Minnan speech, existing resources remain limited, and few provide paired Minnan and Mandarin transcripts at scale. To address these gaps, we introduce WenetSpeech-Min, an open-source corpus comprising around 10,000 hours of Minnan speech collected from diverse online media, with paired Minnan and Mandarin transcripts for every utterance. We further establish an automatic speech recognition (ASR) benchmark covering both Minnan and Mandarin transcripts and a text-to-speech synthesis (TTS) benchmark using Minnan transcripts, with manually verified evaluation sets for both tasks. To assess the effectiveness of the corpus, we train ASR and TTS models on WenetSpeech-Min and compare them with representative systems on the proposed benchmarks. The resulting models outperform the evaluated open-source models on most metrics and achieve competitive performance against commercial systems. We will release the corpus, benchmarks, and models to facilitate reproducible research on Minnan speech technology\footnote{\href{https://aslp-lab.github.io/WenetSpeech-Min-Repo/}{WenetSpeech-Min-Repo webpage}} .

\end{abstract}
\section{Introduction}
\label{sec:intro}

Large-scale and diverse datasets have substantially advanced automatic speech recognition (ASR) and text-to-speech synthesis (TTS) for high-resource languages such as Mandarin and English~\cite{wenetspeech,libritts,Emilia}. Despite these advances, Minnan \footnote{\url{https://en.wikipedia.org/wiki/Southern_Min}} (also known as Southern Min) is widely spoken across Fujian, Taiwan, and Southeast Asia, yet remains under-resourced in speech technology, with limited large-scale public speech corpora.

Minnan speech supports two complementary transcription targets. Dialect transcripts preserve the spoken words and dialect-specific expressions, whereas Mandarin transcripts render the content of the same speech segments in written Mandarin without requiring word-by-word correspondence to the spoken Minnan. The former supports dialect-preserving recognition and linguistic research, while the latter supports Mandarin-output recognition and applications requiring readable written Mandarin. Because the two representations preserve different information and serve different purposes, neither can fully substitute for the other.


Existing Minnan resources address only part of these needs. Mozilla Common Voice~\cite{ardila2020commonvoice} provides a small read-speech corpus, while MinSpeech~\cite{lin2024minspeech} offers thousands of hours of speech with only Mandarin transcripts. YT-THDC~\cite{yang2026tgasr} includes paired dialect and Mandarin transcripts but only 30 hours of domain-specific speech. Thus, large-scale real speech with both transcription targets remains unavailable. For evaluation, Breeze Taigi~\cite{lan2026breeze} introduces ASR and TTS benchmarks, but its evaluation data are not publicly available, limiting reproducible evaluation of new systems. GigaSpeechBench~\cite{tu2026gigaspeechbench} evaluates Minnan speech-to-Mandarin recognition but does not support dialect-transcription ASR. 
Consequently, the central data bottleneck is the absence of large-scale real-world speech with paired dialect and Mandarin transcripts, which is increasingly important as speech models continue to scale. Addressing this bottleneck requires a data construction pipeline. Meanwhile, the absence of public benchmarks for paired-transcript ASR and Minnan TTS prevents fair, reproducible, and comprehensive comparison of existing systems under consistent protocols.

Building on the WenetSpeech series~\cite{wenetspeech,wenetspeech_yue,wenetspeech_chuan,wenetspeech_wu}, we introduce \textit{WenetSpeech-Min}, a publicly available, approximately 10,000-hour multi-source corpus of real-world Minnan speech, with each utterance paired with Minnan and Mandarin transcripts. To construct the corpus, we develop a pipeline for processing heterogeneous media and generating paired transcripts. We further establish public benchmarks for paired-transcript ASR and Minnan TTS using high-quality, manually verified evaluation sets, and systematically assess representative open-source and commercial systems under unified evaluation protocols. Finally, we train ASR and TTS models on WenetSpeech-Min which outperform the evaluated open-source models on most metrics and remain competitive with commercial systems, demonstrating the utility of the corpus.

\section{WenetSpeech-Min}
\label{sec:dataset}
\subsection{Data Construction Pipeline}
\label{ssec:data_pipeline}

\begin{figure*}[t]
  \centering
  \includegraphics[width=0.90\textwidth]{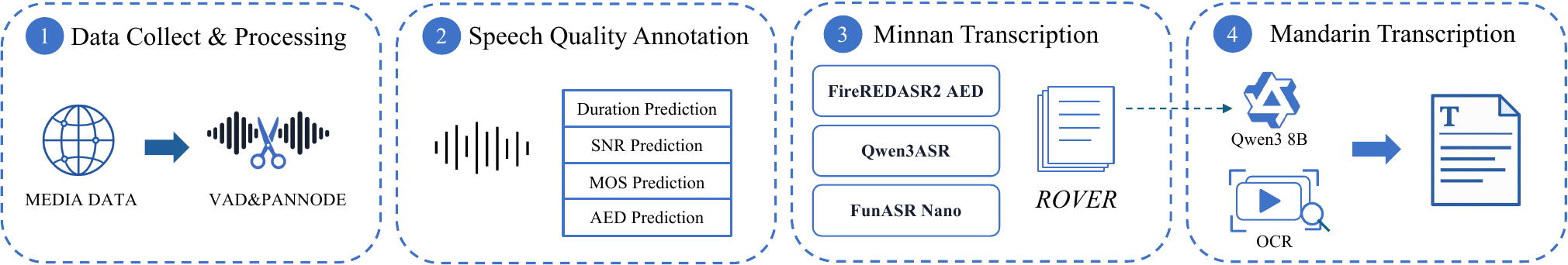}
  \caption{Overview of the WenetSpeech-Min data construction pipeline.}
  \label{fig:data_pipeline}
\end{figure*}

To construct a large-scale Minnan speech corpus from heterogeneous media sources, we develop a unified data construction pipeline, as illustrated in Fig.~\ref{fig:data_pipeline}. The pipeline consists of two main components: audio processing and paired-transcript generation. The first component segments the audio and filters the resulting segments according to their speech content and acoustic quality. The second combines multiple Minnan ASR systems to obtain a dialect transcript and then produces the corresponding Mandarin transcript.

\subsubsection{Audio Processing}
\label{sssec:audio_processing}

We collect Minnan speech from various online media sources and apply voice activity detection (VAD) to segment long-form recordings into utterance-level units. For each segment, we first compute the Signal-to-Noise Ratio (SNR) and the duration. These features are then used to compute a Word-level Virtual Mean Opinion Score (WV-MOS), which characterizes the acoustic quality of the segment. We further apply the DaSheng~\cite{dinkel2024scaling} audio event detection (AED) model to estimate the confidence that a segment contains speech and remove segments below a predefined threshold. Finally, we use pyannote~\cite{bredin2023pyannote} to estimate the number of speakers in each retained segment; this information is stored as segment-level metadata for subsequent corpus analysis.

\subsubsection{Annotation Tool Construction}
\label{sssec:annotation_models}

To support large-scale automatic transcription, we construct dedicated annotation models using manually transcribed Minnan speech. Specifically, 1,600 hours of speech with manually verified Minnan transcripts are used to fine-tune Qwen3-ASR~\cite{qwen3asr}, FireRedASR2-AED~\cite{xu2026fireredasr2s}, and Fun-ASR-Nano~\cite{an2025fun}. These three models subsequently serve as complementary ASR annotators for generating dialect-transcript candidates. A 600-hour subset is further annotated and manually verified with paired Minnan and Mandarin transcripts. Using the paired text from this subset, we fine-tune Qwen3-8B~\cite{qwen3} with LoRA~\cite{hu2022lora} to obtain a specialized Minnan-to-Mandarin translation model for generating Mandarin transcripts.


\subsubsection{Paired-Transcript Generation}
\label{sssec:transcript_generation}

Each retained utterance is associated with a dialect transcript and a Mandarin transcript. To generate the dialect transcript, three ASR annotation models described above produce complementary hypotheses, which are aligned and combined using Recognizer Output Voting Error Reduction (ROVER)~\cite{fiscus1997rover}. The combined result is retained as the dialect transcript.

The Mandarin transcript is obtained through one of two routes. For media with embedded subtitles, we use WeSubtitle\footnote{\url{https://github.com/wenet-e2e/wesubtitle}} to extract the subtitle text through optical character recognition (OCR), and the normalized subtitle text is used directly. Otherwise, the fine-tuned Qwen3-8B translation model converts the ROVER-combined dialect transcript into written Mandarin. Candidate transcripts are subsequently normalized, and consistency checks are applied to each audio--text pair. Samples with unreliable segmentation or low-confidence content are removed or returned for correction.

\begin{figure*}[t]
  \centering
  \begin{minipage}[t]{0.235\textwidth}
    \centering
    \includegraphics[width=\linewidth]{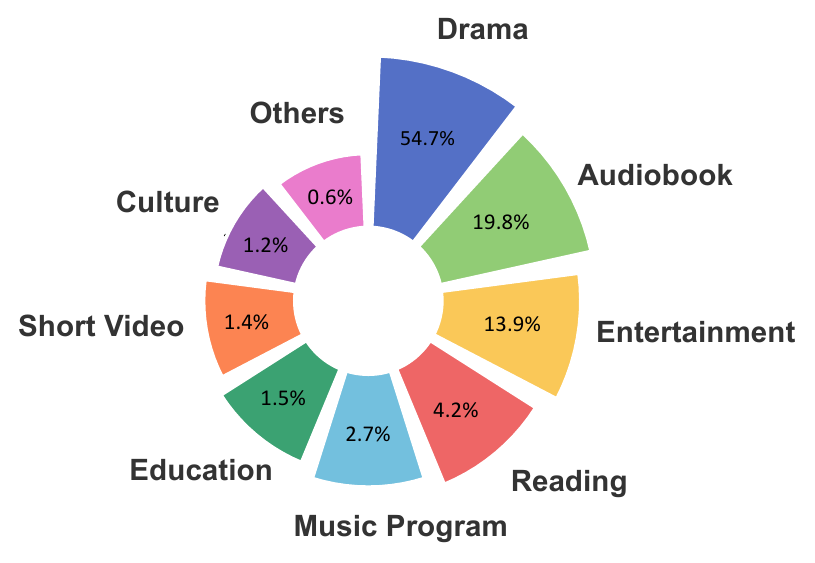}
    \small (a) Content-domain distribution.
  \end{minipage}
  \hfill
  \begin{minipage}[t]{0.235\textwidth}
    \centering
    \includegraphics[width=\linewidth]{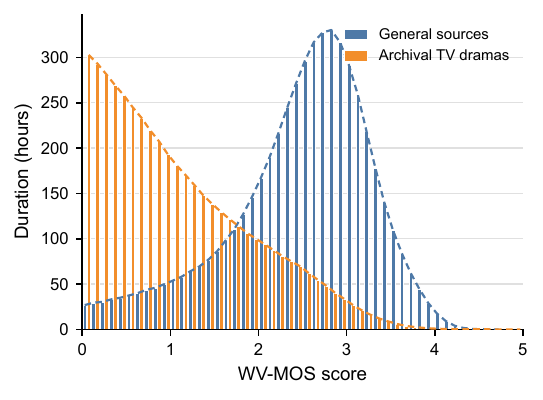}
    \small (b) WV-MOS distribution.
  \end{minipage}
  \hfill
  \begin{minipage}[t]{0.235\textwidth}
    \centering
    \includegraphics[width=\linewidth]{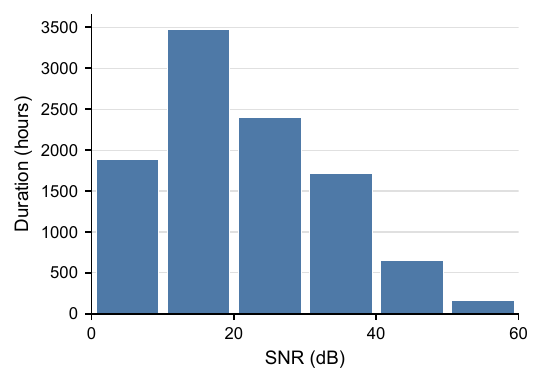}
    \small (c) SNR distribution.
  \end{minipage}
  \hfill
  \begin{minipage}[t]{0.235\textwidth}
    \centering
    \includegraphics[width=\linewidth]{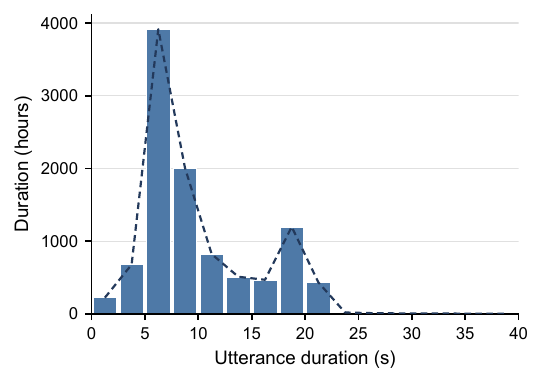}
    \small (d) Duration distribution.
  \end{minipage}
  \caption{Distributions characterizing WenetSpeech-Min: (a) content domains over the sourced hours, (b) WV-MOS by source group, (c) SNR, and (d) utterance duration.}
  \label{fig:dataset_distributions}
\end{figure*}  

\newcommand{\benchmarkresulttables}{%
\begin{table}[!t]
  \caption{Minnan transcription results on WS-Min-Eval-ASR. Lower D-CER is better. Bold indicates the best performance, underlining indicates the second-best performance, and \colorbox{wenetspeechmin}{light green} rows indicate models trained on WenetSpeech-Min.}
  \label{tab:asr_dialect_results}
  \centering
  \small
  \setlength{\tabcolsep}{5pt}
  \begin{tabular}{lc}
    \toprule
    \textbf{Model} & \textbf{D-CER (\%)$\downarrow$} \\
    \midrule
    Hy-ASR-3.0-Preview$^\dagger$ & 32.78 \\
    CN-MultiDialect-ASR & 18.59 \\
    Qwen3-ASR & 34.61 \\
    \rowcolor{wenetspeechmin}
    Qwen3-ASR-WSM-Min & \underline{17.59} \\
    \rowcolor{wenetspeechmin}
    Qwen3-ASR-WSM-Min + internal data & \textbf{15.21} \\
    \midrule
    FireRedASR2-AED & 43.39 \\
    \rowcolor{wenetspeechmin}
    FireRedASR2-AED-WSM & 18.95 \\
    \bottomrule
  \end{tabular}
  \par\smallskip
  \begin{minipage}{\columnwidth}
    \footnotesize $^\dagger$ Results obtained via a commercial API.
  \end{minipage}
\end{table}

\begin{table*}[!t]
  \caption{Mandarin transcription results on WS-Min-Eval-ASR and the external GigaSpeechBench and MinSpeech test sets. Avg. denotes the unweighted mean across the three test sets. Lower M-CER is better, and higher M-BLEU is better. Bold indicates the best performance, underlining indicates the second-best performance, and \colorbox{wenetspeechmin}{light green} rows indicate models trained on WenetSpeech-Min.}
  \label{tab:asr_mandarin_results}
  \centering
  \footnotesize
  \setlength{\tabcolsep}{0.7pt}
  \begin{tabular}{@{}lcccccccc@{}}
    \toprule
    \textbf{Model}
    & \multicolumn{2}{c}{\textbf{WS-Min-Eval-ASR}}
    & \multicolumn{2}{c}{\textbf{GigaSpeechBench}}
    & \multicolumn{2}{c}{\textbf{MinSpeech}}
    & \multicolumn{2}{c}{\textbf{Avg.}} \\
    \cmidrule(r){2-3} \cmidrule(lr){4-5} \cmidrule(lr){6-7} \cmidrule(l){8-9}
    & \textbf{M-BLEU$\uparrow$}
    & \textbf{M-CER (\%) $\downarrow$}
    & \textbf{M-BLEU$\uparrow$}
    & \textbf{M-CER (\%) $\downarrow$}
    & \textbf{M-BLEU$\uparrow$}
    & \textbf{M-CER (\%) $\downarrow$}
    & \textbf{M-BLEU$\uparrow$}
    & \textbf{M-CER (\%) $\downarrow$} \\
    \midrule
    \rowcolor{gray!12}
    \multicolumn{9}{@{}l}{\textbf{Mandarin Transcription}} \\

    FunASR-Realtime$^\dagger$
    & 39.09 & 45.41
    & 53.37 & 32.62
    & 28.22 & 63.23
    & 40.23 & 47.09 \\

    SeedASR2.0$^\dagger$
    & 38.91 & 47.58
    & \underline{55.32} & \underline{32.50}
    & \textbf{68.53} & 26.21
    & \textbf{54.25} & 35.43 \\

    Qwen3-ASR-MinSpeech
    & 20.34 & 60.57
    & 12.39 & 71.26
    & \underline{67.94} & \textbf{23.93}
    & 33.56 & 51.92 \\

    \rowcolor{wenetspeechmin}
    Qwen3-ASR-WSM-Mandarin
    & 42.93 & 42.56
    & 51.20 & 34.01
    & 59.82 & 29.42
    & 51.32 & 35.33 \\

    \rowcolor{wenetspeechmin}
    Qwen3-ASR-WSM-Mandarin + internal data
    & \textbf{49.39} & \textbf{36.81}
    & 47.65 & 36.66
    & 65.69 & \underline{25.68}
    & \underline{54.24} & \textbf{33.05} \\

    \midrule

    \rowcolor{gray!12}
    \multicolumn{9}{@{}l}{\textbf{Paired-Transcript Recognition}} \\

    FireRedASR2-AED
    & 18.05 & 64.40
    & 36.91 & 49.35
    & 13.60 & 85.02
    & 22.85 & 66.26 \\

    \rowcolor{wenetspeechmin}
    FireRedASR2-AED-WSM
    & \underline{46.06} & \underline{39.53}
    & \textbf{57.40} & \textbf{28.61}
    & 57.73 & 31.85
    & 53.73 & \underline{33.33} \\

    \bottomrule
  \end{tabular}
  \par\smallskip
  \begin{minipage}{\textwidth}
    \footnotesize $^\dagger$ Results obtained via commercial APIs.
  \end{minipage}
\end{table*}

\begin{table*}[!t]  
  \caption{TTS results on the WS-Min-Eval-TTS. Lower CER is better, and higher is better for the other metrics. Bold indicates the best performance, underlining indicates the second-best performance, and \colorbox{wenetspeechmin}{light green} rows indicate models trained on WenetSpeech-Min.}
  \label{tab:tts_results}
  \centering
  \footnotesize
  \setlength{\tabcolsep}{4pt}
  \begin{tabular}{lccccc@{\hspace{3pt}}ccccc}
    \toprule
    \textbf{Model}
    & \multicolumn{5}{c}{\textbf{WS-Min-Eval-TTS-Easy}}
    & \multicolumn{5}{c}{\textbf{WS-Min-Eval-TTS-Hard}} \\
    \cmidrule(r){2-6} \cmidrule(l){7-11}
    & \textbf{CER (\%)$\downarrow$}
    & \textbf{SIM$\uparrow$}
    & \textbf{I-MOS$\uparrow$}
    & \textbf{S-MOS$\uparrow$}
    & \textbf{A-MOS$\uparrow$}
    & \textbf{CER (\%)$\downarrow$}
    & \textbf{SIM$\uparrow$}
    & \textbf{I-MOS$\uparrow$}
    & \textbf{S-MOS$\uparrow$}
    & \textbf{A-MOS$\uparrow$} \\
    \midrule
    QwenAudio-3.0-TTS$^\dagger$
    & \textbf{19.90} & 0.672 & \underline{3.88} & \textbf{3.85} & 3.62
    & \textbf{35.09} & \underline{0.685} & \textbf{3.75} & \underline{3.73} & \underline{3.70} \\
    Qwen3TTS-Flash$^{\dagger,\ddagger}$
    & 24.44 & -- & \textbf{3.93} & -- & \textbf{3.93}
    & 38.02 & -- & \textbf{3.75} & -- & \textbf{3.73} \\
    \midrule
    FireRedTTS3
    & 33.76 & \textbf{0.747} & 3.30 & 3.32 & 3.13
    & 55.33 & \textbf{0.731} & 3.15 & 3.39 & 3.04 \\
    VoxCPM2
    & 28.16 & \underline{0.707} & 3.45 & 3.54 & 3.38
    & 40.50 & 0.681 & 3.17 & 3.48 & 3.36 \\
    MERaLiON-TTS
    & 21.25 & 0.622 & 3.83 & 3.73 & \underline{3.65}
    & 37.28 & 0.636 & 3.58 & \underline{3.73} & 3.63 \\
    CosyVoice3
    & 63.69 & 0.654 & 2.13 & 2.13 & 1.49
    & 60.35 & 0.610 & 2.00 & 2.40 & 1.43 \\
    \rowcolor{wenetspeechmin}
    CosyVoice3-WSM
    & \underline{20.26} & 0.678 & 3.85 & \underline{3.75} & \underline{3.65}
    & \underline{35.98} & 0.669 & \underline{3.65} & \textbf{3.85} & \underline{3.70} \\
    \bottomrule
  \end{tabular}
  \par\smallskip
  \begin{minipage}{\textwidth}
    \footnotesize $^\dagger$ Results obtained via commercial APIs. $^\ddagger$ Uses a single fixed speaker; SIM and S-MOS are therefore not evaluated.
  \end{minipage}
\end{table*}
}

\subsection{Dataset}
\label{ssec:statistics}

WenetSpeech-Min is a large-scale Minnan speech corpus with paired dialect and Mandarin transcripts. It contains approximately 10,000 hours of speech collected from multiple online platforms. In the following, we describe the data scale, source distribution, and acoustic quality of the corpus.

\subsubsection{Data Scale and Source Distribution}
\label{sssec:source_distribution}

WenetSpeech-Min contains approximately 10,000 hours of speech, segmented into 5,121,249 utterances, and every retained utterance carries both a dialect transcript and a Mandarin transcript. The utterances vary considerably in length, with a mean duration of 7.24~s and a median falling in the 5--7.5~s interval. The material collected is categorized by content domain. Because approximately 1,500 hours of audio lack source records, the domain distribution is computed over the remaining roughly 8,500 hours with recorded provenance. Fig.~\ref{fig:dataset_distributions}(a) reports the resulting distribution. Drama dominates the corpus, accounting for 54.7\% of the sourced hours, followed by Audiobook (19.8\%) and Entertainment (13.9\%); together these three domains cover more than 88\% of the sourced material, while the remaining hours are spread over a long tail of smaller categories.

\subsubsection{Audio Quality}
\label{sssec:audio_quality_statistics}

We characterize the acoustic conditions of the corpus using WV-MOS, SNR, and duration. Fig.~\ref{fig:dataset_distributions}(b) compares the WV-MOS distributions of the two major source groups. The general-source subset has an approximate duration-weighted mean of 2.47, with 6.9\% of its duration below 1.0, while the archival television subset has a mean of 1.09, with 54.2\% below 1.0. The latter consists mainly of television dramas from the 1990s to the early 2010s. Its lower scores largely reflect the acoustic characteristics inherited from the original recording, broadcast, and archival conditions rather than the degradation introduced by our data construction pipeline. We retain this material because its extended dramatic scenes provide natural multi-speaker interaction, emotional speech, and diverse character-dependent speaking styles. WenetSpeech-Min therefore combines relatively clean speech with challenging legacy-media recordings, favoring broader linguistic and expressive coverage over uniformly high signal quality.

Fig.~\ref{fig:dataset_distributions}(c) shows the SNR distribution in 1-dB intervals for the released utterances. The distribution is largely concentrated between 10 and 40 dB, which represents 73.8\% of the total duration. The duration-weighted mean SNR is 21.3~dB, and the median falls in the 19–20~dB range, indicating that the released corpus provides predominantly moderate-to-high acoustic quality while retaining a substantial range of recording conditions.


Fig.~\ref{fig:dataset_distributions}(d) presents the utterance-duration distribution in 2.5-s intervals. Most utterances are relatively short, although the distribution exhibits a secondary peak around 20~s. This peak mainly originates from source data that had already been segmented into approximately 20-s clips prior to ingestion; we retained these clips without applying an additional round of VAD-based segmentation. Consequently, the corpus combines predominantly short utterances with a smaller set of longer contiguous speech segments. Although the secondary peak reflects the preprocessing of the source data rather than an explicit segmentation target, these retained segments provide complementary training examples with longer temporal context.

\section{Benchmark}
\label{sec:benchmark}
\subsection{Paired-Transcript ASR Benchmark}
\label{ssec:asr_benchmark}

To evaluate both transcription targets on the same audio, we construct WS-Min-Eval-ASR, a six-hour paired-transcript ASR benchmark. Each utterance is paired with two manually verified references: a Minnan transcript and a Mandarin transcript. This paired design enables both transcription targets to be evaluated using the same audio segments and annotation protocol. All systems follow a unified text-normalization and scoring procedure. For Minnan transcription, we report the dialect character error rate (D-CER) against the Minnan reference. For Mandarin transcription, we report the Mandarin Bilingual Evaluation Understudy (M-BLEU) and the Mandarin Character Error Rate (M-CER) against the Mandarin reference. M-BLEU provides a complementary measure of n-gram overlap when valid Mandarin renderings may differ in wording, whereas M-CER measures character-level accuracy.

\subsection{Minnan TTS Benchmark}
\label{ssec:tts_benchmark}

Table~\ref{tab:tts_benchmark_composition} summarizes WS-Min-Eval-TTS, which comprises expert-verified Minnan texts generated by an LLM and refined for naturalness and dialect authenticity. The Easy subset contains 500 everyday utterances, while the Hard subset contains 250 challenging texts: 46 long-form texts of 161--270 characters, 189 repetition-heavy texts with five to seven repetitions, and 15 tongue twisters. Nine reference speakers are selected from an internal Minnan corpus based on acoustic quality and dialect authenticity. For objective evaluation, intelligibility is measured by CER using transcriptions generated by the Qwen3-ASR-WSM-Min model, while speaker similarity (SIM) is calculated as the cosine similarity between WavLM~\cite{seedttseval} embeddings of the synthesized and reference speech. For subjective evaluation, 23 native Minnan-speaking listeners assessed intelligibility (I-MOS), speaker similarity (S-MOS), and accent authenticity (A-MOS) in 40 randomly selected test cases, with 20 drawn from each subset.


\begin{table}[t]
  \caption{Composition of WS-Min-Eval-TTS. ``\#'' denotes the number of utterances.}
  \label{tab:tts_benchmark_composition}
  \centering
  \footnotesize
  \setlength{\tabcolsep}{4pt}
  \begin{tabular}{@{}lr>{\raggedright\arraybackslash}p{0.62\columnwidth}@{}}
    \toprule
    \textbf{Subset} & \textbf{\#} & \textbf{Composition} \\
    \midrule
    Easy & 500 & Everyday utterances \\
    Hard & 250 & Long, repetition, and tongue-twister texts \\
    \bottomrule
  \end{tabular}
\end{table}

\benchmarkresulttables

\section{Experiments}
\label{sec:experiments}
\subsection{Automatic Speech Recognition}
\label{ssec:asr_results}


Existing Minnan ASR systems support different output targets: some produce Minnan transcripts, whereas others produce Mandarin transcripts. We therefore evaluate each system under its supported output setting on WS-Min-Eval-ASR, together with GigaSpeechBench and MinSpeech for Mandarin transcription. The evaluated systems include the open-source Qwen3-ASR~\cite{qwen3asr}, CN-MultiDialect-ASR~\cite{wang2026policy}, and FireRedASR2-AED~\cite{xu2026fireredasr2s}, as well as the commercial Hy-ASR-3.0-Preview\footnote{\href{https://console.cloud.tencent.com/tokenhub/models/detail?modelId=hy-asr-3.0-preview}{Hy-ASR-3.0-Preview webpage}}, FunASR-Realtime\footnote{\href{https://www.qianwenai.com/models/fun-asr-flash-8k-realtime}{FunASR-Realtime webpage}}, and SeedASR 2.0\footnote{\href{https://docs.byteplus.com/zh-CN/docs/byteplusvoice/speechtotextv2}{SeedASR2.0 webpage}}.  For Minnan transcription, we fine-tune Qwen3-ASR on the Minnan transcripts of WenetSpeech-Min, yielding Qwen3-ASR-WSM-Min. For Mandarin transcription, we fine-tune Qwen3-ASR on MinSpeech to obtain Qwen3-ASR-MinSpeech as an architecture-matched baseline, and fine-tune Qwen3-ASR on the combination of MinSpeech and the Mandarin transcripts of WenetSpeech-Min, yielding Qwen3-ASR-WSM-Mandarin. To further explore the model's potential, we additionally train the models on high-quality internal data.  FireRedASR2-AED is fine-tuned on the paired transcripts to support both outputs within a single model, using two learnable task embeddings to condition the decoder with the CTC loss disabled.

Tables~\ref{tab:asr_dialect_results} and~\ref{tab:asr_mandarin_results} reveal several findings. First, adaptation on WenetSpeech-Min yields strong performance for both transcription targets. Qwen3-ASR-WSM-Min outperforms all evaluated existing systems for Minnan transcription. Further training on high-quality internal data reduces its D-CER to 15.21\%, establishing a new state of the art for Minnan-transcript recognition on WS-Min-Eval-ASR. For Mandarin transcription, Qwen3-ASR-WSM-Mandarin with internal-data training achieves the lowest average M-CER of 33.05\% and the second-highest average M-BLEU of 54.24, nearly matching the best commercial result of 54.25, demonstrating performance competitive with commercial systems.
Second, cross-benchmark score variation may partly reflect differences in the mapping between Minnan speech and Mandarin references. Because the two transcripts are not strictly aligned word by word, different benchmarks involve different degrees of translation or paraphrasing. This may help explain why Qwen3-ASR-MinSpeech performs best on MinSpeech but poorly on the other test sets, and why additional internal-data training improves Qwen3-ASR-WSM-Mandarin on two benchmarks but degrades it on GigaSpeechBench. Thus, absolute M-CER and M-BLEU scores are not directly comparable across benchmarks, although they remain useful for comparing ASR systems within each benchmark.
Finally, FireRedASR2-AED-WSM achieves the best results on GigaSpeechBench while remaining competitive on Minnan and Mandarin transcription. This result shows that paired-transcription training can support both transcription targets within a single model while maintaining competitive cross-domain performance.

\subsection{Speech Synthesis}   
\label{ssec:tts_results}

To verify the effectiveness of our data, we fine-tune the open-source model CosyVoice3~\cite{cosyvoice3}. Specifically, we first perform continual pre-training (CPT) on the single-speaker data in WenetSpeech-Min. Building on this, we filter 1,000 hours of high-quality data according to SNR and MOS scores for supervised fine-tuning (SFT). We then filter synthesized samples using CER and SIM to construct preference pairs and apply direct preference optimization (DPO), yielding CosyVoice3-WSM. We compare CosyVoice3-WSM with the open-source CosyVoice3, FireRedTTS3~\cite{fireredtts3}, VoxCPM2~\cite{voxcpm2}, and MERaLiON-TTS\footnote{\href{https://huggingface.co/MERaLiON/MERaLiON-OmniVoice-Hokkien-TTS}{MERaLiON-TTS webpage}} baselines, as well as the commercial Qwen3TTS-Flash\footnote{\href{https://www.qianwenai.com/models/qwen3-tts-flash}{Qwen3TTS-Flash webpage}} and QwenAudio-3.0-TTS\footnote{\href{https://www.qianwenai.com/models/qwen-audio-3.0-tts-flash}{QwenAudio-3.0-TTS webpage}}.

Table~\ref{tab:tts_results} shows that CosyVoice3-WSM substantially outperforms its base model on both subsets. Fine-tuning consistently reduces CER and improves SIM on both the Easy and Hard subsets. Among the open-source systems evaluated, CosyVoice3-WSM achieves the lowest CER, although its SIM remains below FireRedTTS3 and VoxCPM2. The subjective listening results further show that it achieves the highest I-MOS and S-MOS among the open-source systems on both subsets, as well as the highest A-MOS on Hard. The discrepancy between the SIM and S-MOS rankings may reflect the influence of intelligibility on perceptual speaker-similarity judgments: poorly intelligible speech can receive lower S-MOS ratings even when its embedding-based SIM is relatively high. Although Qwen3TTS-Flash is a single-fixed-speaker API and is therefore not evaluated using SIM or S-MOS, it achieves the highest I-MOS and A-MOS on both subsets, demonstrating its strong synthesis capability. QwenAudio-3.0-TTS achieves the lowest CER on both subsets and strong subjective scores, including the highest S-MOS on Easy and a tie for the highest I-MOS on Hard. Overall, CosyVoice3-WSM achieves performance comparable to commercial systems, supporting the effectiveness of WenetSpeech-Min for Minnan TTS adaptation. 

\section{Conclusion}
\label{sec:conclusion}
In this paper, we present WenetSpeech-Min, a large-scale, multi-source Minnan speech corpus containing approximately 10,000 hours of audio, with paired Minnan and Mandarin transcripts. To construct the corpus, we develop a Minnan speech processing pipeline to handle large volumes of data. We also establish a paired-transcript ASR benchmark and a Minnan TTS benchmark, addressing the lack of comprehensive evaluation resources. To validate the utility of the corpus, we train ASR and TTS models on WenetSpeech-Min, which outperform the evaluated open-source baselines on most metrics and are competitive with commercial systems.


\section{Acknowledgments}

We thank NexData for providing high-quality Minnan speech data used in this study.

\section{Compliance with Ethical Standards}

Speech data used in this study were obtained from publicly accessible online media and used solely for research purposes. All reference speakers and listening-test participants provided informed consent, and their responses were anonymized. The study was conducted in accordance with applicable institutional and ethical requirements.



\bibliographystyle{IEEEbib}
\bibliography{strings,refs}

\end{document}